# Lochon Catalyzed D-D Fusion in Deuterated Palladium in the Solid State

K. P. Sinha,[a] and A. Meulenberg[b]

[a] *INSA Honorary Scientist*
*Department of Physics & NMR Research Centre, Indian Institute of Science, Bangalore-12, India*
(*kpsinha@gmail.com*)

[b] *Visiting Scientist, Department of Instrumentation, Indian Institute of Science, Bangalore-12, India*
(*mules333@gmail.com*)

**Abstract**

Lochons (local charged bosons or local electron pairs) can form on $D^+$ to give $D^-$ (bosonic ions) in Palladium Deuteride in the solid state. Such entities will occur at special sites or in a linear channel owing to strong electron-phonon interaction or due to potential inversion on metallic electrodes. These lochons can catalyze $D^-$ - $D^+$ fusion as a consequence of internal conversion leading to the formation of $^4$He plus production of energy ($Q$ = 23.8 MeV) which is carried by the alpha particle and the ejected electron-pair. The reaction rate for this fusion process is calculated.

(**Keywords**: Lochon mediated nuclear reaction, solid state internal conversion, Condensed matter nuclear science)

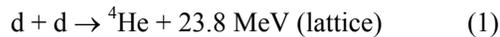

## Introduction

Recent experimental work on palladium, loaded electrolytically with deuterium in the solid state, has given a quantitative correlation between heat and $^4$He production which is consistent with fusion reaction[1].

$$d + d \rightarrow {}^4He + 23.8 \text{ MeV (lattice)} \quad (1)$$

This fusion reaction in the hot plasma state involves electric dipole radiation causing the reversal of the parity of the particle system. Such reactions producing $^3$He or tritium along with neutron or proton are much more probable than those producing $^4$He. Two deuterons (carrying even parity) fusing to produce $^4$He in the first excited state (having even parity) cannot involve electric dipole radiation or an odd-parity particle to reach the ground state (again with even parity).

However, in a solid-state matrix, additional channels are possible which can render the above reaction feasible. These arise from the lattice structure, which harbors quanta of lattice vibrations (phonons) and electron pairs. Owing to strong electron-phonon interaction, there can be negative – U* centers which can have negative charging energy and thus produce local electron pairs in the singlet (S = 0) state [2-5]. These local electron pairs (christened lochons for local charged bosons) can also occur due to potential inversion arising from an electrolytic environment or screening due to metallic electrodes[6-8].

In previous papers[2,3], it has been shown that deuteron pairs reside in crevices, voids or linear channels of the metallic (e.g., Pd) cathodes. The interaction with high frequency modes (optical phonons, polaritons, or surface plasmons) reduces the Coulombic repulsion between two electrons making double occupancy ($n_d$ = 2) more favorable than single occupancy ($n_d$ = 1) of deuterons. As a result D⁻ at certain sites will be more stable. In fact, the electron pair (lochon) behaves like a boson and can produce strong binding between two deuterons by causing resonance exchange[3] – $D^--D^+ \leftrightarrows D^+-D^-$. In the vibration mode, the two nuclei have a finite probability of overcoming the Coulomb barrier leading to their fusion. Denoting the lochon (local pairs – lattice coupled via phonons) by $b = (e_\uparrow e_\downarrow)$, the following reaction can be envisaged,

$$d + d + (b) \rightarrow {}^4He + (b) + Q, \quad (2)$$

with the lochon (b) and ⁴He carrying off the available energy $Q = 23.8$ MeV. The charged boson (b) can excite phonon or plasmon modes and can subsequently split into two fermions electrons ($e_\uparrow$ and $e_\downarrow$). The situation is akin to solid state internal conversion[9].

In what follows, we present the results of calculation based on the above model.

**Calculation of Reaction Constant**

The rate of fusion reaction is expressed as[10]

$$1/\tau_F = A |\phi_i(0)|^2, \quad (3)$$

where $A$ is the nuclear reaction constant (having dimension cm³/sec) and $\phi_i(0)$ is the initial wave function in relative coordinates of fusing particles at zero separation (in fact the interaction distance between the nuclei).

The approximate form of $|\phi_i(0)|^2$ can be taken as

$$|\phi_i(0)|^2 \approx (\mu_{dd}/\mu_b)^{1/2} a_b^{-3} \exp(-G), \quad (4)$$

where $\mu_{dd}$ is the reduced mass of the two dd nuclei, $\mu_b$ is the mass and $a_b$ is the radius of the binding particle (here the pair of electrons in the lochon) and $G$ is the Gamow factor. With the deuterons being confined in a linear channel (one dimensional situation), we have

$$G = (e^*_1 e^*_2 / \hbar v_p), \quad (5)$$

where $e^*_1$ and $e^*_2$ are the screened charges of two deuterons, $\hbar = h/2\pi$, and $v_p$ is their maximum relative parallel velocity.

For solid state internal conversion, the expression for $A$ has been evaluated by Kalman and Keszthelyi[9] by using Weisskopf approximation and pure Coulomb interaction[11]. This can be adapted for the present model, keeping in mind important differences. In the present case, the process involves bound electron pair to free electron pair transition.

We have

$$A = (2/\pi^2) \alpha_f^2 Z_1^2 Z_2^2 (g/V_c) \omega_\alpha I, \quad (6)$$

where $\alpha_f$ is the fine structure constant ($e^2/\hbar c = 1/137$); $Z_1, Z_2$ are the effective charge factors of two electrons in the bound electron pair and two deuterons respectively, $g$ is the degeneracy; $g = 2$, for the present case, $V_c$ is the volume of the interaction cell, $\omega_\alpha = m_\alpha c^2/\hbar$, ($m_\alpha c^2$ being the rest energy of the helium nucleus i.e. the (alpha particle). $I$ is the integral evaluated in reference 9.

$$I = |M_f|^2 (16/25) 128\pi^2 (a_b/Z)^2 K_\alpha^{-1} \quad (7)$$

$|M_f|^2 = (\pi R_o^3/3)$ for $K_\alpha R_o \ll 1$; $R_o$ is the nuclear radius,

$$K_\alpha = [2m_\alpha (Q - 2m_e^* c^2)]^{1/2} / \hbar, \quad (8)$$

$(a_b/Z)$ the effective radius of the bound electron pair on D⁻, $2m_e^* c^2$ being the rest energy of the electron pair, $2m_e^*$ their effective mass inclusive of phonon-interaction effects. This results in the final expression of $A$, $A =$

$$(\frac{2}{\pi^2})\alpha_f^2 Z_1^2 Z_2^2 (\frac{2}{V_c})\omega_\infty \left(\frac{\pi R_o^3}{3K_\alpha}\right) \cdot \left(\frac{16}{25}\right) \cdot 128\pi^2 \left(\frac{a_b}{Z}\right)^2 \quad (9)$$

In the next section, we give the numerical estimate of $|\phi_i(0)|^2$, $A$, and $(1/\tau_F)$.

**Numerical Estimate**

For the Gamow factor $G = e^*_1 e^*_2 / \hbar v_p$, we use the values $e^*_1 = e^*_2 = 0.5e = 2.4 \times 10^{-10}$ esu, $\hbar = 10^{-27}$ ergs sec, and $v_p = 6.5 \times 10^6$ cm sec$^{-1}$ giving $G = 8.75$. Also, $\mu_{dd} = 1.67 \times 10^{-24}$ gms, $\mu_b = 2 \times 10^{-27}$ gms, and $a_b$ (defined in terms of the effective mass $\mu_b$ and effective charge $e_b$ of the lochon) = $\hbar^2 / \mu_b e_b^2$ = $10^{-54}/(2\times10^{-27})(2 \times 4.8 \times 10^{-10})^2 = 5.4 \times 10^{-10}$. Thus, $|\phi_i(0)|^2 \sim (\mu_{dd}/\mu_b)^{1/2} a_b^{-3} \text{Exp}[-G] \sim 2\times10^{29}$ cm$^{-3}$.

For evaluating $A$ (cf. Eq. (9), we use the values, $\alpha_f = (1/137)$, $\omega_\alpha = (m_\alpha c^2/\hbar) = 6 \times 10^{23}$ sec$^{-1}$ with the mass of alpha particle $m_\alpha = 6.64 \times 10^{-24}$ gms, $Z_1 = Z_2 = 2$, $R_o = 5\times10^{-13}$cm, $V_c = (10^{-9})^3 = 10^{-27}$ cm$^3$, $(a_b/Z)^2 = 7.3 \times 10^{-20}$ cm$^2$, with $Z = 2$
$K_\alpha = [2m_\alpha(Q-2m_e^* c^2)]^{1/2}/\hbar = 2.2 \times 10^{13}$ cm$^{-1}$
Using these values, we get $A = 7.5 \times 10^{-19}$ cm$^3$ sec$^{-1}$. Thus, the reaction rate for lochon catalyzed d + d fusion to yield $^4$He (alpha particle) plus 23.8 MeV, energy and ejection of electron pair (eq. 3) is

$$1/\tau_F = A |\phi_i(0)|^2 = 1.5 \times 10^{11} \text{ sec}^{-1}.$$

The above results shows that for lochon mediated reactions leading to the generation of $^4$He, the reaction constant approaches that of muon-catalyzed reactions giving t + p or $^3$He + n processes[10], or the solid state internal conversion processes[9] p + d + (e) → $^3$He + (e).

**Summary**

The above model shows that, by taking into account the formation of lochons, a result of the electron-phonon interaction or the electrolytic environment, adequate screening of the deuterons exists to facilitate d-d fusion. Also, the bosonic character of the lochons permits the formation of the $^4$He nucleus via a new channel. Thus, the present model provides confidence in the low energy nuclear reaction experimental results noted in reference 1.

**Acknowledgement**

This work is supported by Indian National Science Academy; HiPi Consulting, New Market, MD, USA; Science for Humanity Trust, Bangalore, India; and the Science for Humanity Trust, Inc. Tucker, GA, USA.